\documentclass[prb,twocolumn,floats]{revtex4-1}
\usepackage{graphicx}
\newcommand{\bk}{{\bf k}}

\newcommand{\YBCO}{$\mathrm{YBa_2Cu_3O_{6+x}}${ }}
\newcommand{\BSCCO}{$\mathrm{Bi_2Sr_2CaCu_2O_{8+x}}$}

\begin{document}
\title{Microscopic model for the hidden Rashba effect in YBa$_2$Cu$_3$O$_{6+x}$}
\author{W. A. Atkinson}
\affiliation{Department of Physics \& Astronomy, Trent University, Peterborough ON, Canada, K9L 0G2}
\date{\today}
\begin{abstract}
Each unit cell in YBa$_2$Cu$_3$O$_{6+x}$ contains a pair of two-dimensional CuO$_2$ layers.  While the crystal structure is globally inversion symmetric,  the individual layers are not.  This leads, necessarily, to a nonvanishing Rashba spin-orbit coupling (SOC) in the CuO$_2$ layers, with opposite signs of the coupling constant in each layer.   These so-called Rashba bilayers generate hidden spin textures, with a vansishing net spin at each $\bk$-point in the Brillouin zone, but nonvanishing spin textures in each layer separately.    Here, we trace the microscopic origin of the Rashba splitting through the orbital structure of the CuO$_2$ conduction bands,  obtain a generic three-orbital model Hamiltonian, and show that the magnitude of the spin-splitting predicted by density functional theory is $\sim 10$~meV.
\end{abstract}
\maketitle

\section{Introduction}
Metals that have both inversion symmetry and time-reversal symmetry must have doubly degenerate bands;  that is, for any given wavevector $\bk$, there must exist two distinct spin states (``up'' and ``down'') with the same energy.  This degeneracy is broken in noncentrosymmetric crystals (which lack centers of inversion) as a result of spin-orbit coupling, and if the spin-orbit coupling is sufficiently strong the crystal will have observable spin textures that arise from the mismatch in the ``up'' and ``down'' Fermi surfaces.   Even for centrosymmetric crystals, however, so-called {\em hidden} spin textures may appear when individual atoms are located at sites that lack inversion symmetry.\cite{Zhang:2014jw,Yuan:2019eo}  In this case, the degenerate bands will be spatially separated, so that spatially localized spin textures may form.

Rashba bilayer materials are a simple example of how this can happen.\cite{Das:wl,Dong:2015vo,Riley:2014tu,Jones:2014el,Gehlmann:2016vk} These are centrosymmetric three-dimensional materials containing stacks of atomic bilayers.  The monolayers making up each bilayer are inversion pairs, meaning that they can be mapped onto each other through the inversion operation.  They have Rashba coupling constants $\alpha$ and $-\alpha$, respectively, so that the spin textures are opposite in the two monolayers.\cite{Riley:2014tu,Jones:2014el,Gehlmann:2016vk}

Much of the interest in Rashba bilayers is due to the possibility of engineering topological structures,\cite{Das:wl,Dong:2015vo} or to the nontrivial superconducting\cite{Sigrist:2014,Higashi:2016,Yoshida:2012df,Liu:2017} and nematic\cite{Hitomi:2014,Hitomi:2016} phases that can appear.  However, it was recently recognized that many important cuprate high temperature superconductors, notably \YBCO and \BSCCO, are also Rashba bilayers, and that this can have observable consequences.\cite{Harrison:2015jj,Briffa:2016fs} In particular, spin-polarized angle-resolved photoemission (ARPES) experiments in \BSCCO{} have found evidence for $k$-space spin textures that are consistent with Rashba-like physics.\cite{Gotlieb:2018fx}  

In this work, we look at the details of spin-orbit coupling (SOC) in bilayer cuprates.  For concreteness, we take the case of \YBCO (YBCO), which has a simple cystal structure that makes process of tracing the origins of the Rashba SOC straightforward.  In contrast, \BSCCO{} has a complicated $\sqrt{2}\times 5\sqrt{2}$ buckling superstructure\cite{Levin:1994hn} that complicates the analysis.  The work starts from a tight-binding multiorbital model that is informed by available band structure calculations, and through a process of downfolding, arrives at an effective three-band model for the CuO$_2$ band structure.  These calculations establish the roles played by different orbitals, beyond the usual Cu$d_{x^2-y^2}$,   O2$p_x$, and   O3$p_y$ orbitals, and allow us to estimate the size of the coupling constant.  We show that the Rashba bilayer SOC comes from a confluence of the atomic SOC on the Cu sites and the structural ``dimpling'' of the CuO$_2$ planes.

The topic of spin-orbit coupling in cuprate superconductors was widely discussed in the early 1990s.\cite{Coffey:1991jf,Bonesteel:1992wv,Bonesteel:1993ih,Koshibae:1993ba,Koshibae:1994fi,Yildirim:1995jk}  There, the focus was primarily on explaining observed connections between structural phases and magnetic anisotropy\cite{Thio:1988,Nakano:1994}  in  the La-based cuprates, La$_{2-x}$Sr$_x$CuO$_4$ and La$_{2-x}$Ba$_x$CuO$_4$ (although see Ref.~\onlinecite{Bonesteel:1993ih} for a discussion of magnetic phases in  underdoped YBCO).  These are single-layer materials (having a single CuO$_2$ layer per unit cell) that undergo a sequence of tilt-distortions of the CuO$_6$ octahedra as a function of temperature and doping.\cite{Axe:1989}  As in the current work, the combination of atomic SOC on the Cu sites and structural distortions generates spin textures.

 However, there are fundamental differences in the theoretical approach taken in Refs.~\onlinecite{Coffey:1991jf,Bonesteel:1992wv,Bonesteel:1993ih,Koshibae:1993ba,Koshibae:1994fi,Yildirim:1995jk} and that taken here.  The previous work focused on the low-doping regime of the phase diagram, in which the cuprates can be thought of as doped Mott insulators, having localized spins on the Cu$^{2+}$ sites\cite{Johnston:957} that interact weakly with a dilute gas of itinerant holes.  In this case, the dominant electron-electron interaction is believed to be of the Heisenberg spin-spin type, namely $J\sum_{ij} {\bf S}_i\cdot {\bf S}_j$ where ${\bf S}_{i,j}$ are the spins on on the Cu sites $i$ and $j$.  The SOC then generates  small corrections to the interaction of  the  Dzyaloshinskii-Moriya type, $\sum_{ij} {\bf D}_{ij}\cdot ({\bf S}_i\times {\bf S}_j)$.\cite{Coffey:1991jf,Bonesteel:1992wv,Bonesteel:1993ih,Koshibae:1993ba,Koshibae:1994fi,Yildirim:1995jk}    
 In contrast, we focus here on higher dopings, where there is a well-developed Fermi surface and magnetic correlations are weak so that a local-spin picture is inappropriate.  In this limit, the SOC shows up as a correction to the kinetic energy, rather than the electron-electron interactions.

The model is outlined in Sec.~\ref{sec:model}, and it follows closely the tight-binding description of YBCO put forward in Ref.~\onlinecite{Andersen:1995en}, with the key difference that we emphasize spin-orbit physics.  Two main results of the calculations are then given in Sec.~\ref{sec:results}.  The first of these is an estimate of the mangitude $\Delta E_{\bk}\sim 10 \mbox{--}20$~meV of the Rashba spin splitting of the conduction band.  The second is a downfolded 6-band model, comprising three orbital and two spin degrees of freedom, that provides a simple quantitatively meaningful description of SOC in YBCO.  

\section{Model}
\label{sec:model}
\subsection{Overview}
\begin{figure}[tb]
\includegraphics[width=\columnwidth]{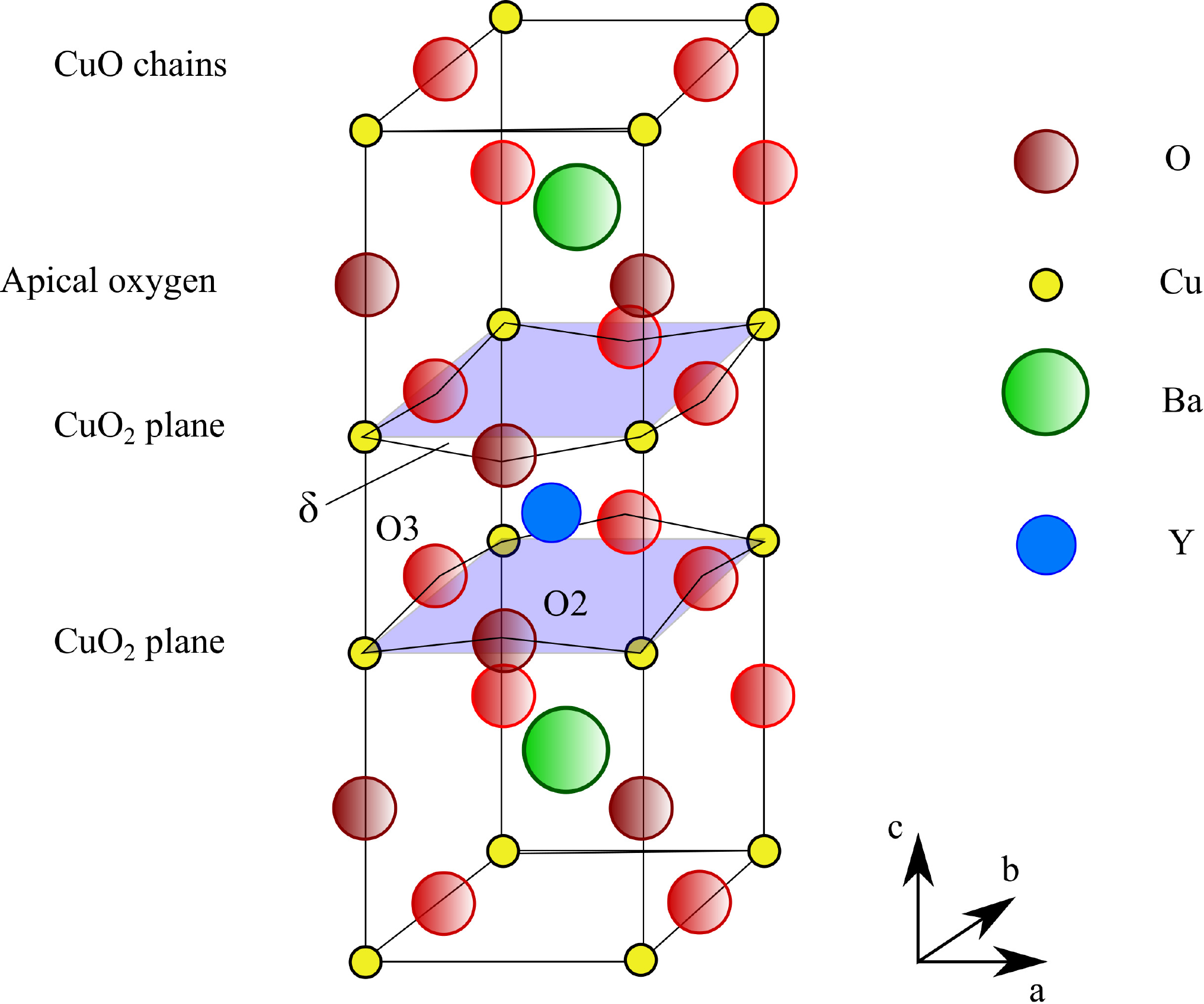}
\caption{Structure of the \YBCO unit cell.  The dimpling angle $\delta$ is shown, as are the locations of the O2 and O3 oxygen sites within the CuO$_2$ planes.}
\label{fig:ybco}
\end{figure}

The YBCO unit cell is shown in Fig.~\ref{fig:ybco}.  As in all of the cuprate high temperature superconductors, the important conduction bands originate in the CuO$_2$ plane layers.  From the figure, it is apparent that all centers of inversion lie outside the CuO$_2$ layers, meaning that inversion symmetry is broken locally within the layers.  From the point of view of the conduction electrons, the largest effects come from the charge imbalance between the Y$^{3+}$ and Ba$^{2+}$ ions that lie on opposite sides of each CuO$_2$ monolayer.  The imbalance creates a local electric field pointing from the Y towards the  Ba ions.  This field directly generates a small Rashba SOC; however, a much more significant effect is to cause dimpling of the CuO$_2$ layers.\cite{Opel:1999gg}    The electric field draws the planar O atoms a distance $\sim 0.25$~\AA{} towards the Y$^{3+}$ layer, so that the Cu-O bond makes a doping-dependent angle $\delta \approx 5\mbox{--}7^\circ$ with the undistorted plane.\cite{Kruger:1997,Kaldis:1997do}  We show in this work that the  Rashba SOC comes  primarily from the combination of dimpling and the atomic spin orbit coupling (ASOC) on the Cu sites.

We can eliminate direct spin-orbit contributions by the electric field as a significant effect:  The strength of the corresponding Rashba constant is $\sim \mu_B v_F E_z/c^2$, where $\mu_B$ is the Bohr magneton, $v_F$ is the Fermi velocity, and $E_z$ the electric field in the CuO$_2$ plane.  Taking $v_F \approx 1.5$~eV\AA{}\cite{Vishik:2010bi} and $E_z \approx 2\mbox{--}4$~V/\AA{}, \cite{Johnston:2010jj}  we get a value of less than 0.01~meV for the constant, which is unobservably small.  This leaves structural distortions as the primary cause of SOC.

\begin{figure}[tb]
\includegraphics[width=\columnwidth]{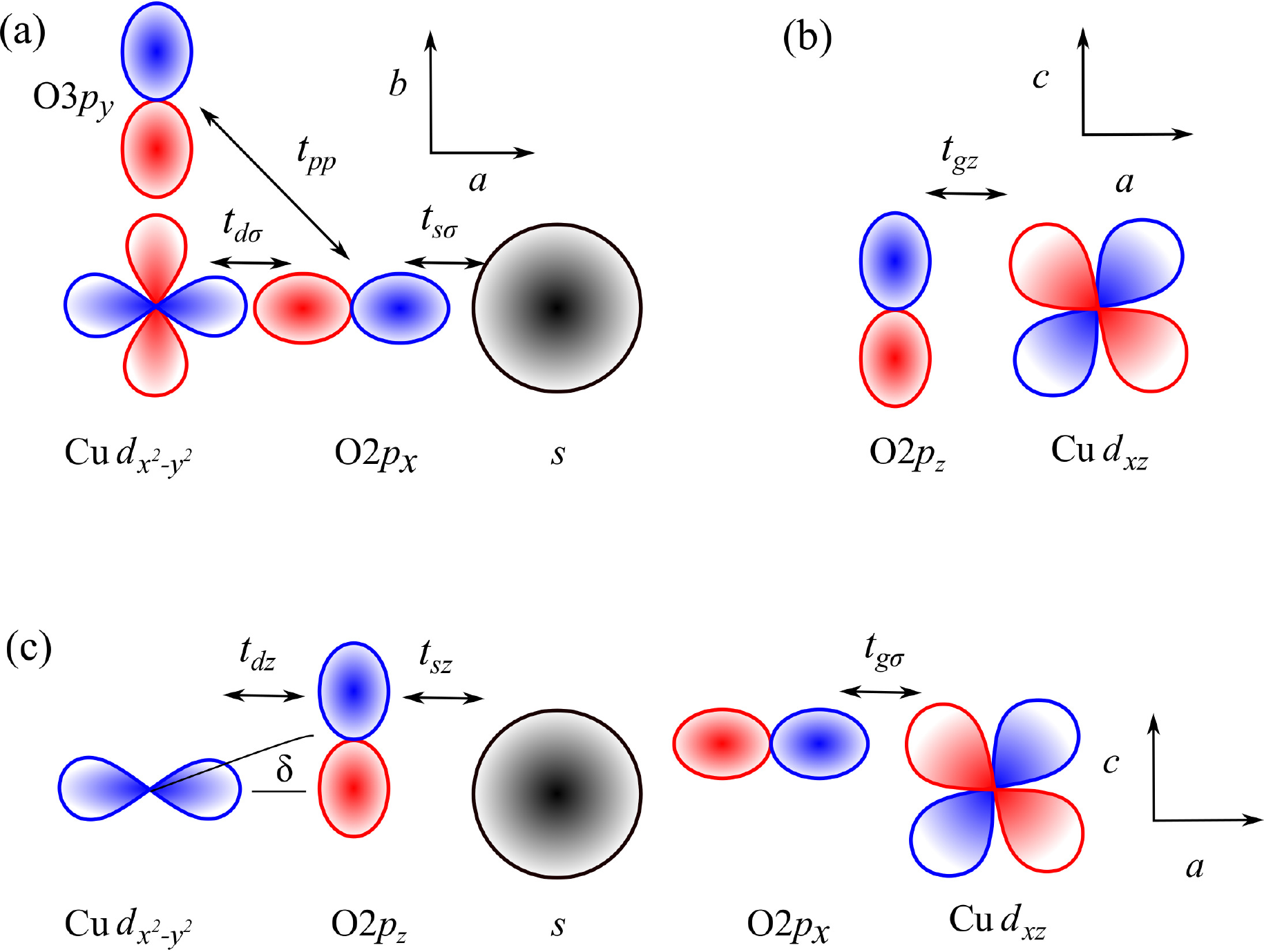}
\caption{The 16-band model, along with relevant hopping matrix elements. (a) Structure of the $\sigma$-bonding block.  The four orbitals shown make the most important contribution to the conduction band.  (b) A total of four orbitals are included in the $\pi$-bonding block:   O2$p_z$,  O3$p_z$, and the two $t_{2g}$ orbitals $d_{xz}$, and $d_{yz}$.  In the absence of dimpling and atomic spin-orbit coupling, the orbital pairs ( O2$p_z$, $d_{xz}$) and ( O3$p_z$, $d_{yz}$) form one-dimensional channels.
(c) Hopping matrix elements that connect the $\sigma$- and $\pi$-blocks.  The dimpling angle $\delta$ is shown, and all of the matrix elements shown in (c) are proportional to $\sin \delta$.   Spin-orbit coupling provides an additional source of mixing between the blocks.}
\label{fig:orbitals}
\end{figure}

We start with a  model for a single CuO$_2$ layer, and then discuss the bilayer in a later section. The model  starts with the four orbitals pictured in Fig.~\ref{fig:orbitals}(a), namely  the Cu$d_{x^2-y^2}$,  O2$p_x$,  O3$p_y$, and a ``Cu$s$'' orbital.   This last orbital represents a mixture of Cu$4s$, Cu$d_{z^2}$, and apical oxygen $p_z$, which hybridize to form a single relevant axially-symmetric orbital.\cite{Pavarini:2001cj}
This orbital has a substantial effect on the Fermi surface shape,\cite{Andersen:1995en,Pavarini:2001cj}  and has been shown to affect the superconducting transition temperature\cite{Pavarini:2001cj,Sakakibara:2010jh} and the structure of the charge density wave in underdoped YBCO.\cite{Atkinson:2015fz,Banerjee:2019vm}    
Together, these orbitals make up the $\sigma$-bonded block of the 8-band Hamiltonian introduced by Andersen {\em et al.} (ALJP) for YBCO.\cite{Andersen:1995en}  In the absence of SOC and dimpling, the $\sigma$-block is sufficient to accurately describe the low-energy portion of the density functional theory (DFT) band structure.\cite{Andersen:1995en}

The second block of orbitals, the so-called $\pi$-block, includes two of the Cu $t_{2g}$ orbitals  ($d_{xz}$ and $d_{yz}$) and the O2$p_z$ and  O3$p_z$  orbitals, which together form a $\pi$-bonded network [Fig.~\ref{fig:orbitals}(b)].\cite{Andersen:1994gp} Without ASOC, the ($d_{xz}$, O2$p_z$) and  ($d_{yz}$, O3$p_z$)  orbital pairs form one-dimensional networks.  ASOC mixes these bands.

The dimpling creates a direct coupling between the $\sigma$- and $\pi$-blocks.  The matrix elements $t_{g\sigma}$, $t_{sz}$, and $t_{dz}$ that couple the two blocks [Fig.~\ref{fig:orbitals}(c)] are proportional to $\sin \delta$ and vanish in the absence of dimpling.  This $\sigma$-$\pi$ mixing was shown in Ref.~\onlinecite{Andersen:1994gp} to significantly affect the band structure near the antinodes [i.e.\ near $\bk = (\pi,0)$ and $(0,\pi)$].

ASOC also directly couples the $\sigma$- and $\pi$-blocks, in this case through a mixing of the $d_{x^2-y^2}$ and $t_{2g}$ orbitals.  The Rashba SOC depends on both $\sigma$-$\pi$ mixing and on ASOC, and the basic mechanism is simple:  A valence electron in a $d_{x^2-y^2}$ orbital mixes with $d_{xz}$ and $d_{yz}$ orbitals through ASOC, and then hops into neighbouring  O2$p_x$ and  O3$p_y$ orbitals; this process couples spin-flips with nearest-neighbor hopping. 

\subsection{Hamiltonian}
\label{sec:Hamiltonian}
 As described in the previous section, we keep eight orbitals which, when spin is included, generate a  $16\times 16$ Hamiltonian.   Conventions for the matrix element signs and phases are described in Appendix~\ref{app:mat}.
Values of the Hamiltonian parameters are primarily obtained from DFT, and are given in Table~\ref{table:params}.  The $\sigma$-blocks have the ordered bases
\begin{eqnarray}
|1 \rangle &=& \{ \begin{array}{ccccc}
  |d\uparrow\rangle, & |x\uparrow\rangle, & |y\uparrow\rangle, & |s\uparrow\rangle
\end{array} \}, \\
| 2 \rangle &=&   \{ \begin{array}{ccccc}
|d\downarrow\rangle, & |x\downarrow\rangle, & |y\downarrow\rangle, & |s\downarrow\rangle  
\end{array} \}.
\end{eqnarray}
Here, we have used the compact notation $d \equiv d_{x^2-y^2}$, $x \equiv \mathrm{O2}_x$, and
$y \equiv \mathrm{O2}_y$.
The $\pi$-blocks have the bases
\begin{eqnarray}
|3\rangle &=& \{ \begin{array}{cccc}
 |xz\uparrow\rangle, & |yz\uparrow\rangle, & |z_2 \uparrow\rangle, & |z_3 \uparrow\rangle
\end{array} \}, \\
|4\rangle &=& \{ \begin{array}{cccc}
 |xz\downarrow\rangle, & |yz\downarrow\rangle, & |z_2 \downarrow\rangle, & |z_3 \downarrow\rangle 
\end{array} \},
\end{eqnarray}
with $xz \equiv d_{xz}$, $z_2 \equiv \mathrm{O2}_z$, etc.
The block structure of the Hamiltonian is then
\begin{equation}
H^\mathrm{16b}_\bk = \begin{array}{c|cccc} 
& |1\rangle & |2\rangle & |3\rangle &|4\rangle \\
\hline
\langle 1| & H^{\sigma\sigma}_{\bk} & 0 & H^{\sigma\pi}_\bk & H^{\xi}_{14} \\
\langle 2| & & H^{\sigma\sigma}_{\bk} & H^{\xi}_{23} & H^{\sigma\pi}_\bk \\
\langle 3| &  &  & H^{\pi\pi}_{\bk}+H^{\xi}_{33} & 0 \\
\langle 4| & & & & H^{\pi\pi}_{\bk}+H^{\xi}_{44}
\end{array}
\label{eq:16band}
\end{equation}
with elements below the diagonal obtained from the Hermiticity of $H_\bk$.

\begin{table}[tb]
\begin{tabular}{ll}
Parameter & Value (eV) \\
\hline
$\epsilon_d$ & 0 \\
$\epsilon_{x}$, $\epsilon_{y}$ & $-0.9$ \\
$\epsilon_s$ & 6.5 \\
$\epsilon_{z}$ & 0.4 \\
$\epsilon_g$\cite{Hozoi:2011ui} & -1.5 \\
\hline 
$t_{d\sigma}$ &  1.5 \\
$t_{s\sigma}$ & 2.3 \\
$t_{pp}$ & 0.0 \\
\hline 
$t_{gz}$ & 0.7 \\
$t_{dz}$ & 0.24 \\
$t_{sz}$ & 0.0 \\
$t_{g\sigma}$ & 0.0\\
$\xi$\cite{Ghiringhelli:2002cq} & 0.11 \\
\hline
\end{tabular}
\caption{Parameter values for YBCO, taken from Ref.~\onlinecite{Andersen:1995en} unless otherwise indicated.}
\label{table:params}
\end{table}

The Hamiltonian $H^{\sigma\sigma}_\bk$  is independent of spin, and is 
 \begin{equation}
 \begin{array}{c|cccc}
& |d\rangle & |x \rangle & |y\rangle & |s\rangle \\
\hline
\langle d| &\epsilon_d & 2 t_{d\sigma} s_x & -2t_{d\sigma} s_y & 0 \\
\langle x| & 2 t_{d\sigma} s_x & \epsilon_x & 4 t_{pp} s_x s_y & 2t_{s\sigma} s_x \\
\langle y| & -2t_{d\sigma} s_y & 4 t_{pp} s_x s_y & \epsilon_y & 2t_{s\sigma} s_y \\
\langle s| & 0 & 2t_{s\sigma} s_x & 2t_{s\sigma} s_y & \epsilon_s 
 \end{array}
 \label{eq:SigmaBlock}
 \end{equation}
where $s_x = \sin (k_x/2)$ and $s_y = \sin( k_y/2)$. $H^{\sigma\sigma}_\bk$ by itself captures most of the structure of the conduction band near the Fermi level.  

We note that, in Eq.~(\ref{eq:SigmaBlock}), the direct hopping $t_{pp}$ is actually not the dominant path connecting neighboring oxygen atoms; rather, indirect hopping through the $s$ orbital is more significant.  This point is easily illustrated.  Because the $s$ orbital sits well above the Fermi level in energy, it may be downfolded to create an effective three-band model.\cite{Atkinson:2015fz}   This process renormalizes the parameters $\epsilon_x$, $\epsilon_y$, and $t_{pp}$, with the first two given by Eqs.~(\ref{eq:tildeex}) and (\ref{eq:tildeey}) in Appendix~\ref{app:downfold}, and the latter being
 \begin{equation}
\tilde t_{pp} = t_{pp} + \frac{t_{sp}^2}{\epsilon_F - \epsilon_s}.
\label{eq:tpp}
 \end{equation} 
The second term on the right of Eq.~(\ref{eq:tpp}) describes the indirect hopping through the $s$ orbital.  Because the overlap between the $s$ and $p$ orbitals is large, this indirect hopping process is  significantly larger than the direct hopping matrix element $t_{pp}$, which we set to zero following Ref.~\onlinecite{Andersen:1994gp}.

The $\pi$-block Hamiltonian $H^{\pi\pi}_\bk$, which does not include ASOC, is also independent of spin and is
\begin{equation}
\begin{array}{c|cccc}
&  |xz\rangle & |yz\rangle & |z_2 \rangle & |z_3 \rangle  \\
\hline
\langle xz |  & \epsilon_g & 0 & -2t_{gz}s_x & 0 \\
\langle yz |  & 0 &\epsilon_g & 0 & -2t_{gz}s_y \\
\langle z_2 |  & -2t_{gz} s_x & 0 & \epsilon_z & 0 \\
\langle z_3 |  & 0 & -2t_{gz} s_y & 0 & \epsilon_z 
\end{array}.
\label{eq:PiBlock}
\end{equation}
The $\sigma$-$\pi$ mixing $H^{\sigma \pi}_\bk$ is
\begin{equation} 
\begin{array}{c|ccccc}
&  |xz\rangle & |yz\rangle & |z_2 \rangle & |z_3 \rangle  \\
\hline
\langle d| & 0 & 0 & -2 it_{dz} c_x & 2it_{dz} c_y \\
\langle x| & 2it_{g\sigma} c_x & 0 & 0 & 0 \\
\langle y| & 0 & 2it_{g\sigma}c_y & 0 & 0 \\
\langle s| & 0 & 0 & -2it_{sz} c_x& -2it_{sz} c_y 
\end{array}.
\end{equation}
Here, 
$c_x = \cos({k_x}/2)$ and $c_y = \cos({k_y}/2)$.

So far, all of these matrices are diagonal in the spin index.  Spin mixing comes from the $d$ orbitals on the  Cu sites.  The ASOC energy takes the simple form $\xi {\bf L}\cdot {\bf S}$ and generates a splitting between low- and high-total angular momentum states of $\Delta E_\mathrm{SO} = (2\ell +1) \xi/2$.  Photoemission experiments\cite{Ghiringhelli:2002cq} have reported $\Delta E_\mathrm{SO} = 280$~meV, giving $\xi = 110$~meV for the Cu $d$ orbitals. 

As described above, we include the $d_{x^2-y^2}$, $d_{xz}$, and $d_{yz}$ orbitals in our Hamiltonian, but have neglected the $d_{xy}$ orbital and have absorbed the $d_{z^2}$ into an effective $s$ orbital.  While it may seem {\it a priori} unjustified to drop either orbital from the explicit basis, given that they mix with the band structure through ASOC,  we have checked numerically that neither orbital has a significant quantitative effect on our final results.  We will discuss this point further in Sec.~\ref{sec:results6}.

Keeping all five orbitals for later discussion, the ASOC Hamiltonian for the block of states  
$\begin{array}{cccccc} |d\uparrow \rangle, & |z^2 \uparrow \rangle, & |xy\uparrow\rangle, &|xz\downarrow\rangle, & |yz\downarrow \rangle \end{array}$ is
\begin{equation}
\xi [{\bf L}\cdot {\bf S}] = \frac{\xi}{2} \left [
\begin{array}{ccccc}
0 & 0 & -2i & 1 & i \\
0 & 0 & 0 & -\sqrt{3} & \sqrt{3}i \\
2i & 0 & 0 &-i & 1 \\
1 & -\sqrt{3} & i & 0 & i \\
-i & -\sqrt{3}i & 1 & -i & 0
\end{array}
\right ].
\label{eq:LdotS1}
\end{equation}
The nonzero matrix elements of $H^\xi_{14}$ and $H^\xi_{44}$ in Eq.~(\ref{eq:16band}) can be obtained from Eq.~(\ref{eq:LdotS1}).
Similarly,  the nonzero matrix elements of $H^\xi_{23}$ and $H^\xi_{33}$ are contained in $\xi {\bf L}\cdot {\bf S}$ for the block of states
$\begin{array}{cccccc} |d\downarrow \rangle, & |z^2 \downarrow \rangle, & |xy\downarrow\rangle, &|xz\uparrow\rangle, & |yz\uparrow \rangle \end{array}$ 
\begin{equation}
\xi [{\bf L}\cdot {\bf S}] = \frac{\xi}{2} \left [
\begin{array}{ccccc}
0 & 0 & 2i & -1 & i \\
0 & 0 & 0 & \sqrt{3} & \sqrt{3}i \\
-2i & 0 & 0 &-i  & -1 \\
-1 & \sqrt{3} & i & 0 & -i \\
-i & -\sqrt{3}i & -1 & i & 0 
\end{array}
\right ].
\label{eq:LdotS2}
\end{equation}
Equations (\ref{eq:SigmaBlock}) and (\ref{eq:PiBlock})--(\ref{eq:LdotS2}) provide all the matrix elements required for Eq.~(\ref{eq:16band}).
\begin{figure}
\includegraphics[width=\columnwidth]{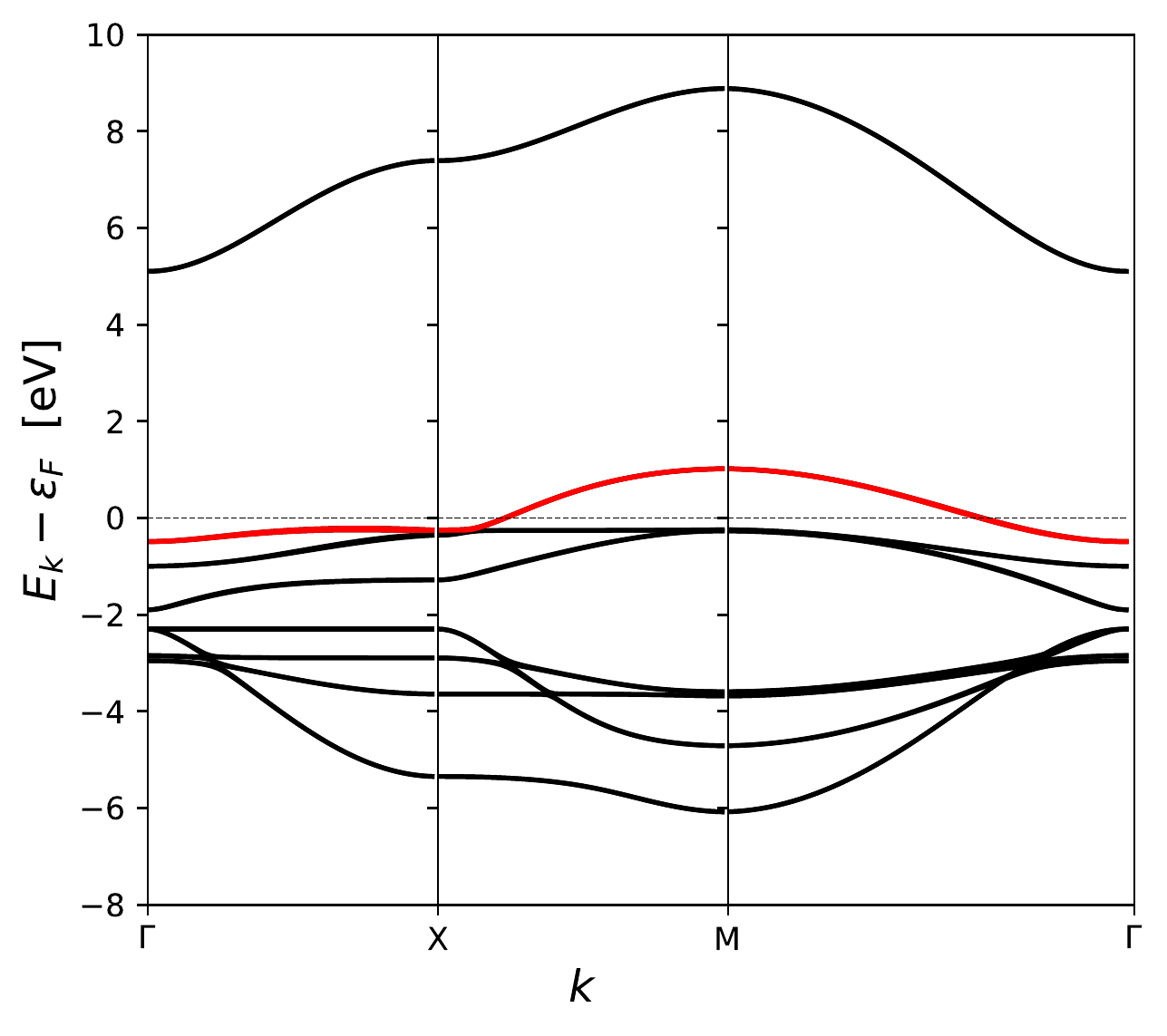}
\caption{Band structure predicted by the 16-band model, obtained from the eigenvalues of Eq.~(\ref{eq:16band}).  The two Rashba-split conduction bands that cross the Fermi surface are in red.  Note that the splitting is less than the thickness of the line.}
\label{fig:spaghetti}
\end{figure}
The band structure obtained from this model is shown in Fig.~\ref{fig:spaghetti}.  It is very similar to that discussed in Ref.~\onlinecite{Andersen:1995en}, on which it is based.

\section{Results}
\label{sec:results}
\subsection{16-band model}
\begin{figure}
\includegraphics[width=\columnwidth]{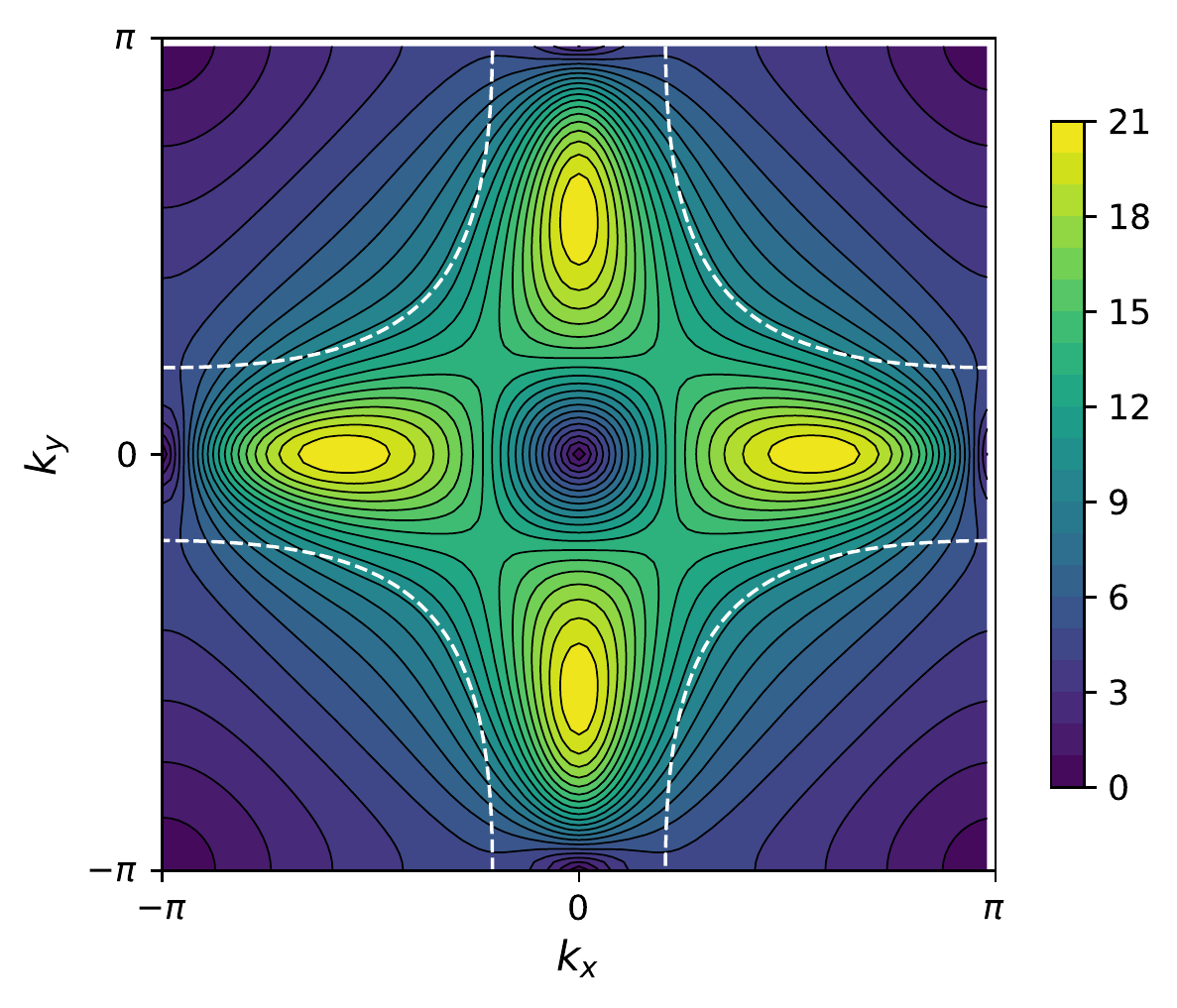}
\caption{Energy difference $\Delta E_\bk$ between the two spin-split bands that cross the Fermi surface.  These bands are obtained by numerically diagonalizing the Hamiltonian, Eq.~(\ref{eq:16band}).  The white dashed line shows the Fermi surfaces for a hole filling of $p=0.10$ holes per unit cell, corresponding to a Fermi energy $\epsilon_F = 2.07$~eV.  Note that the Fermi surfaces for the two bands are  indistinguishable in this figure. The color scale is in meV.}
\label{fig:16band}
\end{figure}

Despite the apparent complexity of the model in Eq.~(\ref{eq:16band}), the Fermi surface structure is simple:  there is a single pair of spin-orbit split bands with energies $E_{\bk\pm}$ that cross the Fermi energy (Fig.~\ref{fig:spaghetti}).  The two bands have opposite windings of the spin (helicities) around their respective Fermi surfaces, and the mismatch in the two Fermi surfaces leads to the formation of spin textures in the CuO$_2$ layer.  It is important to note that the helicities of the two CuO$_2$ layers making up a bilayer are opposite, so that the net spin-polarization (summed over layers) vanishes, as required by global inversion symmetry.

In the absence of SOC, the two bands are degenerate, and to obtain a quantitative measure of the spin-orbit coupling, we calculate their splitting  
\begin{equation}
\Delta E_\bk = E_{\bk+} - E_{\bk-}.
\end{equation}
The results of this calculation are shown in Fig.~\ref{fig:16band}, along with the Fermi surface corresponding to a hole filling of $p=0.10$ holes per unit cell.  (Note that the separation between the two Fermi surfaces is less than the thickness of the dashed line.)  At this filling, the splitting is $\Delta E_\bk \approx 10$~meV along most of the Fermi surface, but drops to near zero at the Brillouin zone boundary.  This value is comparable to that estimated by Harrison {\em et al.}\cite{Harrison:2015jj} and  Briffa {\em et al.}\cite{Briffa:2016fs} by modeling quantum oscillation experiments.  Since $\Delta E_\bk$ is independent of the Fermi energy, the spin splitting at different dopings can be obtained by shifting the Fermi surface in Fig.~\ref{fig:16band}.

\begin{figure}
\includegraphics[width=\columnwidth]{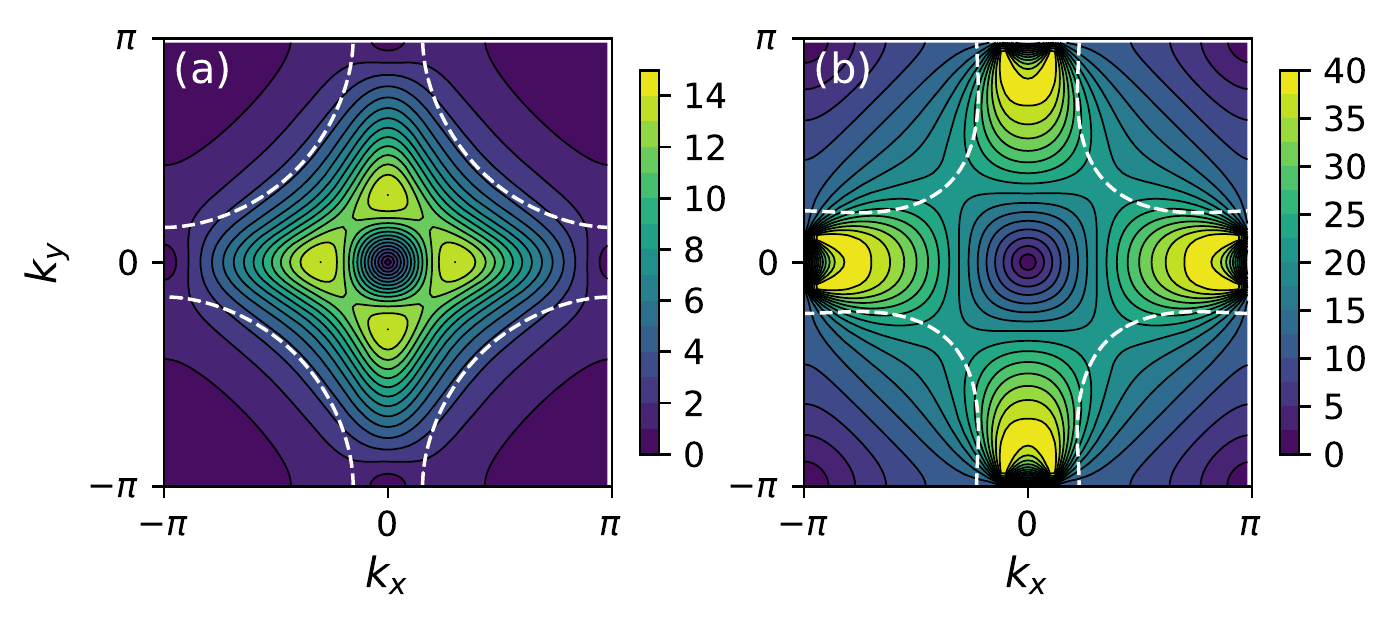}
\caption{Examples showing the range of variation of the spin splitting across different models (see text).  (a) Results for a model in which the Cu$s$ orbital has been removed, with $t_{sp} = 0$ and $t_{pp}=-1.2$~eV.  (b) Results for a model with a reduced hybridization between the Cu$d_{x^2-y^2}$ and oxygen O2$p_x$ and  O3$p_y$ orbitals, with $t_{d\sigma} = 1.0$~eV.} 
\label{fig:compare16}
\end{figure}

Figure~\ref{fig:16band} is one of the two main results of this work.   The parameters underlying this figure are primarily taken from DFT calculations; however, there are well-known discrepancies between DFT and the experimental dispersion that arise because of strong electronic correlations not included in the DFT energy functionals. It is therefore reasonable to ask whether the spin-splitting results are robust.  

We find that for modest perturbations  of the parameters ($\lesssim 50\%$ of the parameter's value), the $k$-space structure of the spin splitting is qualitatively similar to what is shown in Fig.~\ref{fig:16band}, but that the energy scale can change by a factor of up to 2.  To illustrate this point, Fig.~\ref{fig:compare16} shows two of the more extreme cases we explored.

Figure~\ref{fig:compare16}(a) shows the spin splitting for a case in which the  ``Cu$s$'' orbital has been removed by setting $t_{sp}=0$.  To partially compensate for the missing orbital, we set $t_{pp} =-1.2$~eV, which according to Eq.~(\ref{eq:tpp}) should give the same total effective oxygen-oxygen hopping matrix element as in the original calculation with nonzero $t_{sp}$. The resultant Fermi surface has less curvature than the original model, but the dramatic difference is in the spin-splitting along the Fermi surface, which is reduced by a factor of 2 from the original calculation.  

As a second, more physically plausible, example the spin splitting is shown in Fig.~\ref{fig:compare16}(b) for the case in which the hopping matrix element $t_{d\sigma}$ between the $d_{x^2-y^2}$ and O2$p_{x}$/O3$p_y$  orbitals is reduced from 1.5~eV to 1.0~eV.   This leads to a narrowing of the conduction band, as might be expected from strong correlations.  In this case, there is a modest change in the Fermi surface shape compared to Fig.~\ref{fig:16band}, but the spin splitting along the Fermi surface is approximately doubled. 

In summary, numerical calculations suggest that for physically plausible parameter sets the spin-splitting is of order $\Delta E_\bk = 10\mbox{--}20$~meV, and is relatively constant along the Fermi surface, except near the Brillouin zone boundary where it drops to near zero.

\subsection{6-band model}
\label{sec:results6}
To obtain a useful tight-binding characterization of the cuprates, we downfold the 16-band model to an effective 6-band model.  This model includes three orbital degrees of freedom (Cu$d_{x^2-y^2}$, O2$p_x$, and O3$p_y$) and two spin degrees of freedom.  The details of the downfolding process are left to  Appendix~\ref{app:downfold}, and here we give the final results.  It is important to note that while the downfolding procedure gives the correct Fermi surface, it does not give the correct effective band mass, and consequently overestimates the Rashba spin-orbit coupling constant.  We show below that this is a minor issue for the default model parameters, with the predicted spin-splitting being quite close to that of the 16-band model; however, for other model sets the discrepancy can be large.  The value of the downfolding procedure is therefore primarily qualitative; from it, we obtain the generic structure of the Rashba Hamiltonian, along with an understanding of where the different Rashba matrix elements come from.

We first consider the diagonal blocks of the 6-band Hamiltonian,  $H^\mathrm{6b}_\bk$.  The spin-up block is
\begin{eqnarray}
H^\mathrm{6b}_{\bk,\uparrow\uparrow} = 
 \begin{array}{c|ccc} 
& |d\uparrow \rangle & |x\uparrow\rangle   & |y\uparrow\rangle \\
\hline
\langle d\uparrow| & \overline \epsilon_d & 2\overline t^\uparrow_{dpx}s_x & -2\overline t^\uparrow_{dpy}s_y  \\
\langle x\uparrow|  & 2\overline t^\uparrow_{dpx}s_x &  \overline \epsilon_{x} & 4 \overline t^\uparrow_{pp,\bk}  \\
\langle y\uparrow| & -2\overline t^\uparrow_{dpy}s_y  &  4 \overline t^{\uparrow\ast}_{pp,\bk} &  \overline \epsilon_{y} 
\end{array}
\label{eq:6band_diag}
\end{eqnarray}
This has the same structure as the usual three-band model, but with renormalized (and $\bk$-dependent) parameters.  These parameters are given by Eqs.~(\ref{eq:ed})--(\ref{eq:tppk}) in  Appendix~\ref{app:downfold}.  The spin-dependence of the hopping parameters comes from processes in which an electron hops from a $d_{x^2-y^2}$ into one of the $t_{2g}$ orbitals through a nearest-neighbor oxygen, then moves into a different $t_{2g}$ orbital by ASOC, and finally hops back onto an oxygen site.  This is clearly a high-order process, and we have checked that it is much smaller than other spin-dependent terms in the Hamiltonian.

The equations for the matrix elements of $H^\mathrm{6b}_{\bk,\uparrow\uparrow} $ given in Appendix~\ref{app:downfold} are complicated, and for most purposes  a much simpler approximation should be sufficient.  Namely, one may use the parameters from the downfolded four-band $\sigma$-block, Eq.~(\ref{eq:SigmaBlock}).  For this simpler case, one has the renormalized parameters
\begin{eqnarray}
\overline \epsilon_x &=& \epsilon_x + 4 t_{pp}^\mathrm{i} s_x^2 \\
\overline \epsilon_y &=& \epsilon_y + 4 t_{pp}^\mathrm{i} s_y^2 \\
\overline t_{pp} &=& t_{pp} + t_{pp}^\mathrm{i}
\end{eqnarray}
with $s_{x} = \sin(k_x/2)$, $s_{y} = \sin(k_y/2)$, and where 
 the indirect hopping is $t_{pp}^\mathrm{i} = 4t_{sp}^2/(\epsilon_F - \epsilon_s) \approx -1.2$~eV, and all other parameters in Eq.~(\ref{eq:6band_diag}) equal to their bare values.

Next, we turn to the off-diagonal blocks of $H^\mathrm{6b}_{\bk}$ .  To linear order in the ASOC, the ``up-down'' block has the form,
\begin{eqnarray}
H^\mathrm{6b}_{\bk, \uparrow\downarrow} =  \begin{array}{c|ccc} 
& |d\downarrow \rangle & |x\downarrow\rangle   & |y\downarrow\rangle \\
\hline
\langle d\uparrow| & \overline t_{dd\uparrow\downarrow} & \overline t_{dx\uparrow\downarrow} & \overline t_{dy\uparrow\downarrow} \\
\langle x\uparrow|  & \overline t_{xd\uparrow\downarrow} & 0 & 0  \\
\langle y\uparrow| &  \overline t_{yd\uparrow\downarrow}& 0 & 0 
\end{array} 
\label{eq:6band_OD}
\end{eqnarray}
with 
\begin{eqnarray}
\overline t_{dd,\uparrow\downarrow} &=&  -\xi \tilde t_{dg}\left[ \frac{c_y s_y}{\epsilon_F - \tilde \epsilon_{gy}} 
+ i\frac{c_x s_x}{\epsilon_F - \tilde \epsilon_{gx} } \right ] 
\label{eq:tddbar} \\
\overline t_{dx\uparrow\downarrow} &=& \overline t_{xd\uparrow\downarrow} = \frac{ -i\xi t_{g\sigma}  }{\epsilon_F - \tilde \epsilon_{gx}} c_x 
\label{eq:tdx} \\
\overline t_{dy\uparrow\downarrow} &=& \overline t_{yd\uparrow\downarrow} =  \frac{ \xi t_{g\sigma} }{\epsilon_F - \tilde \epsilon_{gy}} c_y,
\label{eq:tdy}
\end{eqnarray}
and  $\tilde \epsilon_{gx}$, $\tilde \epsilon_{gy}$, and $\tilde t_{dg}$ given by Eqs.~(\ref{eq:egx}), (\ref{eq:egy}), and (\ref{eq:tdg}).
The ``down-up'' block is the Hermitian conjugate of Eq.~(\ref{eq:6band_OD}).

Equation~(\ref{eq:6band_OD}) shows that there are two types of terms that are linear in the spin-orbit parameter $\xi$.  The first involves a spin-flip of an electron within the $d_{x^2-y^2}$ orbital; this is actually a multi-step process, in which an electron migrates from the $d_{x^2-y^2}$ orbital to one of the $d_{xz}$ or $d_{yz}$ orbitals by ASOC, and then tunnels back into the $d_{x^2-y^2}$ via one of the planar oxygen $p_z$ states.  The matrix elment for this process is given by Eq.~(\ref{eq:tddbar}).  Noting that $c_a s_a \equiv \cos(k_a/2)\sin(k_a/2)  = \frac 12 \sin k_a$ for $a=x,y$,  Eq.~(\ref{eq:tddbar}) can be usefully approximated as
\begin{equation}
\overline t_{dd,\uparrow\downarrow} = -\alpha_{dd} (\sin k_y + i\sin k_x) 
\end{equation}
which has the Rashba form.

The second type of spin-orbit term is a spin-flip hopping process between the $d_{x^2-y^2}$ and one of the oxygen orbitals.  This process is also multi-step:  a $d_{x^2-y^2}$ electron moves into a $d_{xz}$ or $d_{yz}$ orbital through ASOC, and then hops into a neighboring O2$p_x$ or O3$p_y$ orbital through the matrix element $t_{g\sigma}$.  The matrix elements for this process are given by Eqs.~(\ref{eq:tdx}) and (\ref{eq:tdy}), and  can be usefully approximated as 
\begin{eqnarray}
\overline t_{dx\uparrow\downarrow} &=& -i\alpha_{dp} \cos\frac{k_x}2, \\
\overline t_{dy\uparrow\downarrow} &=& \alpha_{dp} \cos\frac{k_y}2. 
\end{eqnarray}
One can show [Appendix~\ref{app:symmetries}] that these off-diagonal terms have the form expected for Rashba SOC, namely, they break inversion symmetry, are time reversal invariant, and are even under fourfold rotations.

We note that to linear order in the ASOC parameter $\xi$, the $d_{xy}$ orbital does not contribute to the Rashba SOC. This is because the matrix elements coupling this orbital to the  O2$p_x$ and  O3$p_y$ orbitals vanish by symmetry, even with dimpling, so that the only hopping processes with Cu$d_{x^2-y^2}$ and oxygen orbitals as initial and final states must be of order $\xi^2$.  Furthermore, because the spin-orientation is unchanged in these processes, the $d_{xy}$ orbital actually appears only in the renormalization of the diagonal spin-up and spin-down blocks of the 6-band Hamiltonian, and not in the off-diagonal spin-flip blocks.  This justifies its neglect in the original Hamiltonian.

The role of the $d_{z^2}$ orbital, which  we have ignored up to this point, can be established by a similar line of reasoning.  This orbital does not couple directly to the $d_{x^2-y^2}$ orbital via ASOC [see Eqs.~(\ref{eq:LdotS1}) and (\ref{eq:LdotS2})], and it's largest contribution to spin-orbit effects will therefore be as a conduit for oxygen-oxygen hopping.  As an example, the effective matrix element for spin-flip processes in which the electron starts and finishes on an O2$p_x$ orbital is
\begin{equation}
\overline t_{xx\uparrow \downarrow} \approx -\xi \frac{4 t_{z^2\sigma} t_{g\sigma}}{(\epsilon_F-\epsilon_g)^2}
c_x s_x. 
\end{equation}
Here, the electron starting in an O2$p_x$ orbital hops into a neighboring $d_{z^2}$ orbital, moves into the $d_{xz}$ orbital via ASOC, and then hops into one of the two neighboring O2$p_x$ orbitals.  The first of these processes has matrix element $t_{z^2\sigma}$;  
Ref.~\onlinecite{Andersen:1994gp} suggests that it is small for YBCO, and can be neglected; however, if this is not the case, then one must further add the matrix elements
\begin{equation}
-\alpha_{pp} \times \left \{
\begin{array}{c|cc} 
& |x\downarrow\rangle   & |y\downarrow\rangle \\
\hline
\langle x\uparrow| & c_x s_x & i c_y s_x \\
\langle y\uparrow| & c_x s_y & i c_y s_y 
\end{array}\right \},
\end{equation}
to Eq.~(\ref{eq:6band_OD}), with $\alpha_{pp} = 4 \xi t_{z^2\sigma} t_{g\sigma}/(\epsilon_F-\epsilon_g)^2$.  Because $\alpha_{pp}$ comes from higher order hopping processes than $\alpha_{dp}$, one can exect the latter to be larger.

\begin{figure}
\includegraphics[width=\columnwidth]{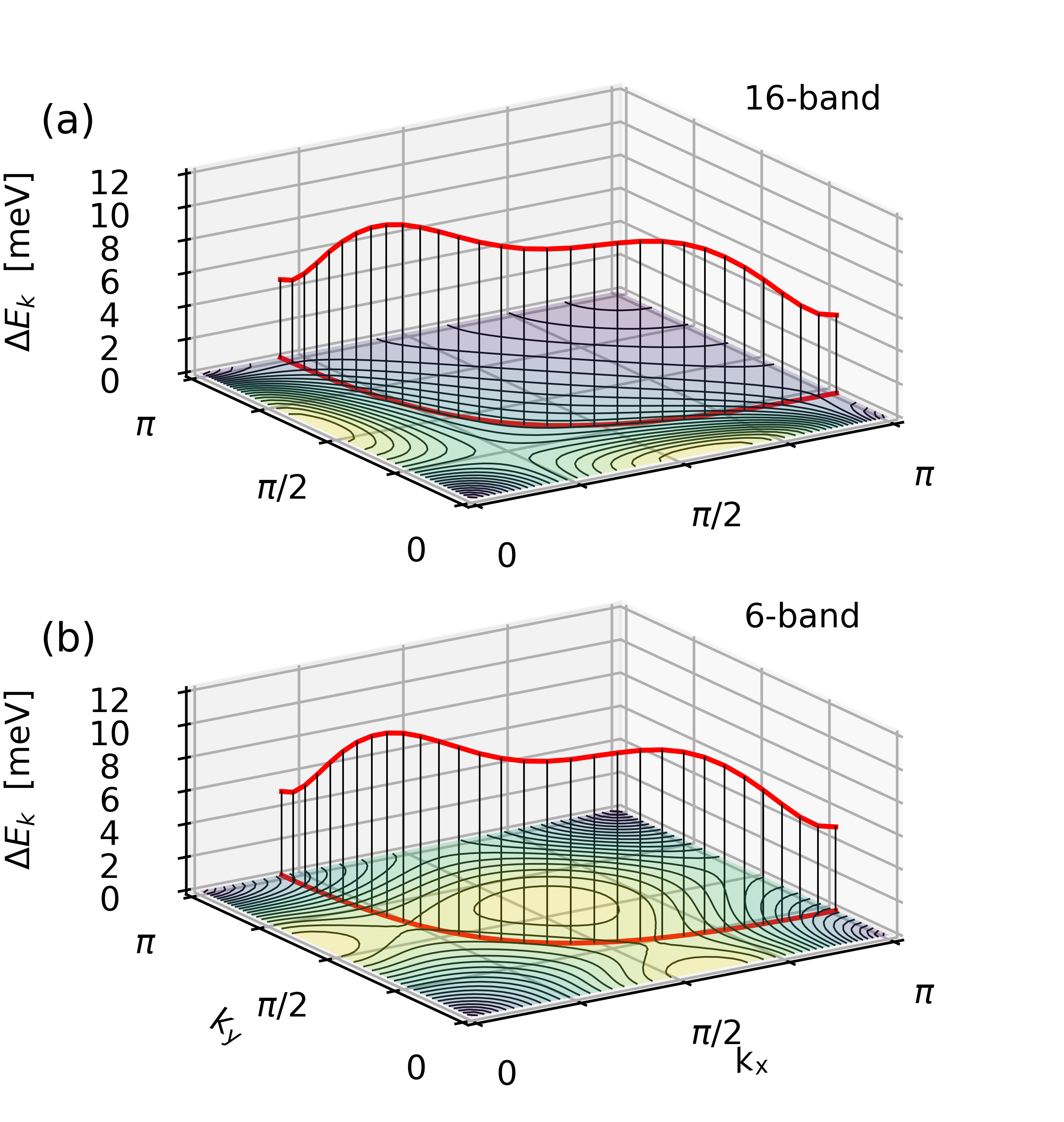}
\caption{Comparison of the spin splittings for (a) the 16-band model and (b) 6-band model, using the same parameters as in Fig.~\ref{fig:16band}.  Contours show the spin splitting throughout one quadrant of the Brillouin zone, while the curves show the value of the spin splitting along the Fermi surface.  The two curves differ only by a few percent and show the accuracy of the 6-band model near the Fermi surface.}
\label{fig:6v8}
\end{figure}

The quality of the 6-band downfolding is illustrated in Fig.~\ref{fig:6v8}, where the spin-splitting is shown in one quadrant of the Brillouin zone for both models. (The 6-band calculations use the full expressions for the model parameters, from Appendix~\ref{app:downfold}, rather than the simplified parameters discussed above.) The contour plots show the spin-splitting throughout the quadrant, while the curves show the spin-splitting along the Fermi surface.  Two features of these figures are notable:   the contour maps are very different, yet the plots along the Fermi surface are nearly the same, differing by only $\sim 10\%$.  As mentioned above, this discrepancy can be primarily attributed to the failure of the downfolding process to correctly include mass-renormalization effects, and can be larger for other choices of parameter. 
Ultimately, the value of the downfolded Hamiltonian is that it provides a generic model Hamiltonian for YBCO.

\subsection{The Bilayer}
(Much of the discussion in this section can be found elsewhere,\cite{Harrison:2015jj,Briffa:2016fs,Maharaj:2016} and is repeated here for completeness.)
The extent to which spin-orbit effects are relevant to the CuO$_2$ bilayer depends on the relative strengths of the Rashba coefficient $\alpha$ and the interlayer hopping $t_{\perp,\bk}$.  For simplicity, we consider a one-band model for the CuO$_2$ monolayers, such that the total bilayer Hamiltonian has the form
 $\hat H = \sum_{\bk} \Psi^\dagger(\bk) h_0(\bk) \Psi(\bk)$ with 
\begin{equation}
h_0(\bk) = \left [ \begin{array}{cc} 
\epsilon_\bk + ( {\bf g}_\bk \times {\bf \hat z} )\cdot \sigma  & t_{\perp,\bk} \\ 
t_{\perp,\bk} &  \epsilon_\bk -( {\bf g}_\bk \times {\bf \hat z})\cdot \sigma 
\end{array} \right],
\end{equation}
where  ${\bf g}_\bk = \alpha(\sin k_x, \sin k_y)$ is the Rashba SOC
and $\Psi(\bk) = [ c_{1\bk\uparrow},\, c_{1\bk\downarrow},\, c_{2\bk\uparrow},\, c_{2\bk\downarrow}]^T$
is an array of fermion annihilation operators for the two monolayers and spin states.

The Hamiltonian can be diagonalized analytically, and one obtains two doubly-degenerate bands, with energies $\xi_{\bk\pm} = \epsilon_\bk \pm \sqrt{t_{\perp,\bk}^2 + g_\bk^2}$, where $g_{\bk} = |{\bf g}_\bk|$.  This expression shows that the energy splitting between the two bands depends on both the interlayer hopping and the Rashba coupling; furthermore, the character of the bands depends on which of the two terms is larger.  This is most easily seen from the spin texture at the Fermi surface.  It is straightforward to show that the $x$-component of the spin at a specific $\bk$ value in layer 1 is
\begin{eqnarray}
\langle \hat S_{1\bk, x} \rangle &=& \frac 12\langle c^\dagger_{1\bk\uparrow} c_{1\bk\downarrow} +
c^\dagger_{1\bk\downarrow} c_{1\bk\uparrow} \rangle \nonumber \\
&=&  \left [  f(\xi_{\bk+}) - f(\xi_{\bk-}) \right ] \frac{g_\bk \cos \theta_\bk }{\sqrt{t_{\perp,\bk}^2 + g_\bk^2}} 
\label{eq:s1x}
\end{eqnarray}
where $f(\xi_{\bk\pm})$ are the Fermi-Dirac functions for the two bands, and $\cot \theta_\bk = \sin k_y / \sin k_x$.   $\langle \hat S_{1\bk, y} \rangle$ is found by making the substitution $\cos \theta_\bk \rightarrow -\sin\theta_\bk$.\cite{note:inversion} 
The key point of this expression is that the moment in layer 1 is  proportional to
\begin{equation}
\frac{g_\bk }{\sqrt{t_{\perp,\bk}^2 + g_\bk^2}}, 
\end{equation}
which gives us a quantitative measure of the importance of Rashba SOC.  Importantly, when $t_{\perp,\bk} \gg g_\bk$, the interlayer coupling is dominant and quenches SOC.

 It is frequently argued that interlayer coupling in cuprate superconductors has the form $t_{\perp,\bk} = t_\perp (\cos k_x - \cos k_y)^2$; for such a form, the bilayer splitting vanishes along the Brillouin zone diagonals ($k_x = \pm k_y$), at which points any residual splitting must be due to SOC.  This is especially important for cuprate superconductors because the superconducting order parameter vanishes along these directions.  As a result,  low-energy quasiparticles lie along the Brillouin zone diagonals and should be heavily influenced by SOC.
 
 However, this is not the case for YBCO.  Early DFT calculations found that for fully oxygenated YBa$_2$Cu$_3$O$_7$, there is a large splitting along the diagonal directions, of order 100~meV. \cite{Andersen:1995en}   ARPES  experiments\cite{Okawa:2009cl} on overdoped YBCO confirm this large splitting, and tight-binding fits to the data further find that $t_{\perp,\bk}$ is nearly independent of $\bk$ everywhere along the Fermi surface.\cite{Pasanai:2010ky}   All of this strongly suggests that SOC is largely irrelevant in highly oxygenated YBCO.

The story is different in underdoped YBCO.   ARPES experiments by Fournier {\em et al.}\cite{Fournier:2010kk} found that the bilayer splitting collapses  in underdoped YBCO, for hole dopings $p<0.15$. At these doping levels, it  appears plausible that the predicted Rashba spin splitting, $\Delta E_{\bk } \approx 2\alpha = 10\mbox{--}20$~meV, could be the dominant energy scale.

\subsection{Relevance to underdoped \YBCO}
The doping range over which we expect the Rashba SOC to be appreciable corresponds to the region in which both superconductivity and charge density wave order are found.  Here, we ask whether SOC can influence either of these phases.

First, we note that the energy difference between the spin-split bands, $\Delta E_\bk$, corresponds to a Fermi surface splitting of  $\Delta k = \Delta E_\bk/v_F \sim 10^{-2}$~\AA.  The two spin-orbit split Fermi surfaces have opposite helicity, and the predominantly singlet pairing occurs between electrons in the same band.  The slight difference in Fermi wavevector for the two helicities generates a weak triplet component to the superconducting gap.\cite{Sigrist:2009gu} By directly solving the gap equation with parameters appropriate for YBCO, we obtain a triplet component that is approximately 1\% of the dominant singlet component.\cite{Atkinson:2020}  This is unlikely to be important.

The spin-splitting is more likely to be important for the charge-ordered phase.  The splitting of the Fermi surface leads to a multiplicity of nesting wavevectors $q$, differing by amounts $\Delta q \sim \Delta k$; the corresponding length scale $\ell_\mathrm{SOC} = \frac{2\pi}{\Delta q} \sim 600$~\AA{} is the distance over which two charge density waves with different nesting wavevectors will dephase and come into phase again; half that distance is the dephasing length, and because the charge density correlation length is comparable to this length scale,\cite{Chang:2016gz} one can expect the energetics of density wave formation to be affected by SOC.  In particular, we speculate that the dephasing length might set an upper bound for the CDW correlation length.

Finally, we note that simulations have shown that, because the unit cell is polar, there is substantial band-bending at YBCO surfaces.\cite{Pasanai:2010ky}   This breaks the degeneracy of the CuO$_2$ bilayers near the surface, and should reveal the Rashba SOC.  As pointed out in Ref.~\onlinecite{Yuan:2019eo}, one of the key characteristics of hidden-Rashba systems is a strong sensitivity to even weak perturbations that break inversion symmetry.  Thus, we anticipate that the size of the surface Rashba effect should be considerably larger than one might expect from the size of the potential gradient at the surface.  This conjecture remains to be tested, however.

\section{Conclusions}
We have traced the origins of the hidden Rashba spin-splitting in \YBCO via a multiorbital model obtained primarily from density functional theory calculations.  Through a process of downfolding, we obtained a three-orbital model for the the low-energy physics and showed that the spin splitting of the conduction band is $\sim 10$--20~meV.   Finally, we discussed the extent which this effect might be plausibly observed, and argued that it could be have observable consequences in moderately underdoped \YBCO with hole dopings $p < 0.15$.

\section*{Acknowledgments}
We thank Arno Kampf for helpful coversations.  We acknowledge funding support from the Natural Sciences and Engineering Research Council (NSERC) of Canada Discovery Grant program.

\appendix
\section{Matrix element conventions}
\label{app:mat}
The hopping matrix elements in Sec.~\ref{sec:Hamiltonian} are defined with the following conventions.
\begin{enumerate}
\item The matrix elements are assumed to be real, implying that the orbital wavefunctions are  real.  
\item The sign of the hopping matrix element  between two adjacent orbitals is determined by the signs of the closest lobes of each orbital.  If the signs are the same, then the matrix element is positive; otherwise, the matrix element is negative.  Thus, in Fig.~\ref{fig:orbitals}(a), the matrix element between the Cu$d_{x^2-y^2}$ orbital and the  O2$p_x$ to its right is $-t_{d\sigma}$, while the bond to the left has matrix element $+t_{d\sigma}$.   The purpose of this convention is to keep track of the relative signs of different bonds, and the value of $t_{d\sigma}$ itself may be negative or positive depending on the structure of the bond.
\item After Fourier transforming to $k$-space, the piece of the Hamiltonian connecting Cu$d_{x^2-y^2}$ and  O2$p_x$ states has the form $-2it_{d\sigma} \sin \frac{k_x}2 d^\dagger_{\bk \sigma} p_{x \bk \sigma}$, where $d_{\bk\sigma}$ and $p_{x\bk \sigma}$ are fermion operators for the two orbitals.  We make the conventional gauge transformation $p_{x\bk\sigma} \rightarrow i p_{x\bk\sigma}$ to eliminate the factor of $i$.  Similar transformations are made for the O2$p_z$, O3$p_y$, and O3$p_z$ fermion operators.  These gauge transformations eliminate factors of $i$ in $H^{\sigma\sigma}_\bk$ and $H^{\pi\pi}_\bk$, but introduce them in $H^{\sigma\pi}_\bk$. 
\end{enumerate}

As a result of the  gauge transformation, the time reversal of $p_{x\bk\uparrow}$ is $-p_{x-\bk\downarrow}$, i.e.\ there is an addtional factor of $-1$ under time reversal.  This holds also for the other oxygen $p$ orbitals.

\section{Downfolding the Hamiltonian}
\label{app:downfold}
Given a Hamiltonian with the block structure
 \begin{equation}
{H} = \left [ \begin{array}{cc} H_{11} & H_{12} \\ H_{21} & H_{22} \end{array} \right ]
 \end{equation}
 the Green's function for the first block is
 \begin{eqnarray}
 G_{11}(\omega) &=& \left  [( \omega - {H}) ^{-1} \right ]_{11}
 \nonumber \\
& =& \left [ \omega - H_{11} - H_{12} (\omega - H_{22})^{-1}H_{21} \right ]^{-1}.
 \end{eqnarray}
 By inspection, we see that $G_{11}(\omega)$  could equally be obtained from the effective Hamiltonian
 \begin{equation}
 H_{11}^\mathrm{eff} = H_{11} + H_{12} (\omega - H_{22})^{-1} H_{21}.
 \end{equation}
 To obtain a low-energy theory, we set $\omega = \epsilon_F$.  
 
 By this process, we can downfold our $16\times 16$ Hamiltonian into a $6\times 6$ Hamiltonian that includes only $d_{x^2-y^2}$,  O2$p_x$ and O$3_y$ orbitals plus spin degrees of freedom.  To make the process physically transparent, we do this in two steps.  In the first step, we downfold the  O2$p_z$,  O3$p_z$, and $s$ orbitals, and in the second we downfold the $d_{xz}$ and $d_{yz}$ orbitals.

\subsubsection{First downfolding}
Following the first downfolding, we have a $10\times 10$ Hamiltonian with the block structure
\begin{equation}
\tilde H_\bk = 
\left [ \begin{array}{cc} \tilde H_{\bk, 11} & \tilde H_{\bk, 12}  \\  \tilde H_{\bk, 21} & \tilde H_{\bk, 22} 
\end{array} \right ]. 
\end{equation}
The first block, $\tilde H_{\bk,11}$ has the usual structure for three-band models of cuprates, with spin-up matrix elements 
\begin{eqnarray}
 \begin{array}{c|ccc} 
& |d\uparrow \rangle & |x\uparrow\rangle   & |y\uparrow\rangle \\
\hline
\langle d\uparrow| & \tilde \epsilon_d & 2t_{d\sigma}s_x & -2t_{d\sigma}s_y  \\
\langle x\uparrow|  & 2t_{d\sigma}s_x &  \tilde \epsilon_{x} & 4\tilde t_{pp} s_x s_y \\
\langle y\uparrow| & -2t_{d\sigma}s_y  &  4\tilde t_{pp} s_x s_y &  \tilde \epsilon_{y} 
\end{array}
\end{eqnarray}
and an identical set of matrix elements for the spin-down matrix elements.  The parameters are renormalized, however, with $\tilde t_{pp}$ given by Eq.~(\ref{eq:tpp}), and 
\begin{eqnarray}
\tilde \epsilon_d &=& \epsilon_d + \frac{4t_{dz}^2(c_x^2 + c_y^2)}{\epsilon_F - \epsilon_z} \\
\tilde \epsilon_x &=& \epsilon_x + \frac{4t_{sp}^2 s_x^2}{\epsilon_F - \epsilon_s} \label{eq:tildeex} \\
\tilde \epsilon_y &=& \epsilon_y + \frac{4t_{sp}^2 s_y^2}{\epsilon_F - \epsilon_s}. \label{eq:tildeey}
\end{eqnarray}
In this basis, the three diagonal ``orbital energies'' are $k$-dependent, which leads to a significant re-weighting of the oxygen and copper contributions to the Fermi surface.  This re-weighting was found to have a profound effect on charge density-wave formation,\cite{Atkinson:2015fz,Banerjee:2019vm} and in general should not be neglected.

The $\tilde H_{\bk,22}$ block is 
\begin{eqnarray}
 \begin{array}{c|cccc} 
& |xz\downarrow\rangle   & |yz\downarrow\rangle &
  |xz\uparrow\rangle   & |yz\uparrow\rangle \\
\hline
\langle xz\downarrow| & \tilde \epsilon_{gx} & \frac{i\xi}2 &0 &0 \\
\langle yz\downarrow| & -\frac{i\xi}2 & \tilde \epsilon_{gy} &0 &0\\
\langle xz\uparrow| &0&0 & \tilde \epsilon_{gx} & -\frac{i\xi}2  \\
\langle yz\uparrow| &0&0 & \frac{i\xi}2 & \tilde \epsilon_{gy}
\end{array}
\end{eqnarray}
with 
\begin{eqnarray}
\tilde \epsilon_{gx} &=& \epsilon_g + \frac{4 t_{gz}^2 s_x^2}{\epsilon_F - \epsilon_z} 
\label{eq:egx} \\
\tilde \epsilon_{gy} &=& \epsilon_g + \frac{4 t_{gz}^2 s_y^2}{\epsilon_F - \epsilon_z} 
\label{eq:egy} 
\end{eqnarray}
Again, this has the same structure as before the downfolding, but with renormalized orbital energies.

Finally, the $\tilde H_{\bk,12}$ block is
\begin{eqnarray}
 \begin{array}{c|cccc} 
&  |xz\downarrow\rangle   & |yz\downarrow\rangle &
 |xz\uparrow\rangle   & |yz\uparrow\rangle \\
\hline
\langle d\uparrow |& \frac{\xi}2 & \frac{i\xi}2 &  i \tilde t_{dg}c_xs_x & -i \tilde t_{dg}c_ys_y  \\
\langle x\uparrow |&0&0&2it_{g\sigma}c_x&0 \\
\langle y\uparrow |&0&0&0&2it_{g\sigma}c_y \\
\langle d\downarrow |&i \tilde t_{dg}c_xs_x & -i \tilde t_{dg}c_ys_y & -\frac{\xi}2 & \frac{i\xi}2 \\ 
\langle x\downarrow |&2it_{g\sigma}c_x&0&0&0 \\
\langle y\downarrow |&0&2it_{g\sigma}c_y&0&0
\end{array} \nonumber \\
\end{eqnarray}
where 
\begin{equation}
\tilde t_{dg} = \frac{4t_{dz} t_{gz}}{\epsilon_F-\epsilon_z},
\label{eq:tdg}
\end{equation}
is an effective matrix element for hopping between the $d_{x^2-y^2}$ and $d_{xz/yz}$ orbitals through the oxygen $p_z$ intermediate state.  The $\tilde H_{\bk, 21}$ block is the Hermitian conjugate of $\tilde H_{\bk, 12}$.  

\subsubsection{Second downfolding}
The effective 6-band Hamiltonian is
\begin{equation}
 H^\mathrm{6b}_\bk  = \tilde H_{\bk, 11} + \tilde H_{\bk,12} (\epsilon_F - \tilde H_{\bk, 22})^{-1} 
 \tilde H_{\bk, 21}.
\end{equation}
For the sake of transparency, we work under the assumption that $\xi$ is a small parameter and keep only terms to first order in $\xi$.  Then
\begin{widetext}
\begin{equation}
H^\mathrm{6b}_\bk = 
 \begin{array}{c|cccccc} 
& |d\uparrow \rangle & |x\uparrow\rangle   & |y\uparrow\rangle 
& |d\downarrow \rangle & |x\downarrow\rangle   & |y\downarrow\rangle \\
\hline
\langle d\uparrow| & \overline \epsilon_d & 2\overline t^\uparrow_{dpx}s_x & -2\overline t^\uparrow_{dpy}s_y   & \overline t_{dd\uparrow\downarrow} & \overline t_{dx\uparrow\downarrow} & \overline t_{dy\uparrow\downarrow} \\
\langle x\uparrow|  & 2\overline t^\uparrow_{dpx}s_x &  \overline \epsilon_{x} & 4 \overline t^\uparrow_{pp,\bk}  & \overline t_{xd\uparrow\downarrow} & 0 & 0 \\
\langle y\uparrow| & -2\overline t^\uparrow_{dpy}s_y  &  4 \overline t^{\uparrow\ast}_{pp,\bk} &  \overline \epsilon_{y} & \overline t_{yd\uparrow\downarrow} & 0 & 0 \\
\langle d\downarrow| & \overline t_{dd\downarrow\uparrow} & \overline t_{dx\downarrow\uparrow} & \overline t_{dy\downarrow\uparrow} & \overline \epsilon_d & 2\overline t^\downarrow_{dpx}s_x & -2\overline t^\downarrow_{dpy}s_y  \\
\langle x\downarrow|&\overline t_{xd\downarrow\uparrow} & 0 & 0 & 2\overline t^\downarrow_{dpx}s_x &  \overline \epsilon_{x} & 4 \overline t^\downarrow_{pp,\bk}  \\
\langle y\downarrow| &\overline t_{yd\downarrow\uparrow} & 0 & 0& -2\overline t^\downarrow_{dpy}s_y  &  4 \overline t^{\downarrow\ast}_{pp,\bk} &  \overline \epsilon_{y} 
\end{array}
\label{eq:mat6b}
\end{equation}
\end{widetext}
with parameters,
\begin{eqnarray}
\overline \epsilon_d &=& \tilde \epsilon_d + \frac{\tilde t_{dg}^2 c_x^2s_x^2}{\epsilon_F - \tilde \epsilon_{gx}}
+ \frac{\tilde t_{dg}^2 c_y^2s_y^2}{\epsilon_F - \tilde \epsilon_{gy}} 
\label{eq:ed} \\
\overline \epsilon_x &=& \tilde \epsilon_x + \frac{4 t_{g\sigma}^2 c_x^2}{\epsilon_F - \tilde \epsilon_{gx}} \\
\overline \epsilon_y &=& \tilde \epsilon_y + \frac{4 t_{g\sigma}^2 c_y^2}{\epsilon_F - \tilde \epsilon_{gy}} \\
\overline t^ s_{dpx} &=&t_{d\sigma}+ \frac{t_{g\sigma}\tilde  t_{dg} c_x^2}{\epsilon_F - \tilde \epsilon_{gx}}  \nonumber \\ &&
- s \frac{i\xi}{2} \frac{ t_{g\sigma} \tilde t_{dg} c_x c_y}{(\epsilon_F - \tilde \epsilon_{gx})(\epsilon_F - \tilde \epsilon_{gy})} \\
\overline t^{ s}_{dpy} &=& t_{d\sigma} + \frac{t_{g\sigma}\tilde  t_{dg} c_y^2}{\epsilon_F - \tilde \epsilon_{gy}}  \nonumber \\ &&
+ s \frac{i\xi}{2} \frac{ t_{g\sigma} \tilde t_{dg} c_x c_y}{(\epsilon_F - \tilde \epsilon_{gx})(\epsilon_F - \tilde \epsilon_{gy})} \\
\overline t^ s_{pp,\bk} &=&  \tilde t_{pp}s_xs_y -  s \frac{i\xi}{2} \frac{t_{g\sigma}^2  c_x c_y}
{(\epsilon_F- \tilde \epsilon_{gx})(\epsilon_F-\tilde \epsilon_{gy})},
\label{eq:tppk}
\end{eqnarray}
and the spin-flip terms  
\begin{eqnarray}
\overline t_{dd,\uparrow\downarrow} &=& \overline t_{dd,\downarrow\uparrow}^\ast = -\xi \tilde t_{dg}\left[ \frac{c_y s_y}{\epsilon_F - \tilde \epsilon_{gy}} 
+ i\frac{c_x s_x}{\epsilon_F - \tilde \epsilon_{gx} } \right ] \\
\overline t_{dx\uparrow\downarrow} &=& \overline t_{xd\uparrow\downarrow} 
=\overline t_{dx\downarrow\uparrow}^\ast = \overline t_{xd\downarrow\uparrow}^\ast 
= \frac{ -i\xi t_{g\sigma}  }{\epsilon_F - \tilde \epsilon_{gx}} c_x 
\label{eq:tdxud} \\
\overline t_{dy\uparrow\downarrow} &=& \overline t_{yd\uparrow\downarrow} = \overline t_{dy\downarrow\uparrow}^\ast = \overline t_{yd\downarrow\uparrow}^\ast = \frac{ \xi t_{g\sigma} }{\epsilon_F - \tilde \epsilon_{gy}} c_y.
\label{eq:tdyud}
\end{eqnarray}

\section{Symmetries}
\label{app:symmetries}
To illustrate the symmetries of $H^\mathrm{6b}_\bk$, we consider two examples:  time inversion and fourfold rotations.  We focus explicitly on the spin-flip blocks of the Hamiltonian, as these are the new to this work.

Under the time reversal operator $\Theta$, the fermion annihilation operators $d_{\bk s}$ and $p_{ a,\bk s}$ transform as
\begin{equation}
\Theta^{-1} d_{\bk  s} \Theta =  s d_{-\bk\, - s}; \qquad \Theta^{-1} p_{ a, \bk  s} \Theta = - s p_{ a, -\bk\, - s}.
\end{equation}
In this equation $ s=\pm$ represents the two spin states and $ a = x,y$ represents the oxygen $p_x$ and $p_y$ orbitals.  Note that there is an extra factor of $-1$ in the equation for $p_{ a, \bk  s}$; this comes from an implicit factor of $i$ in the definiton of the fermion operators $p_{ a,\bk s}$ that was discussed in Appendix~\ref{app:mat}.  

To show invariance under time reversal, we group terms from $H^\mathrm{6b}_\bk$.  
For example, the combination of terms
\begin{equation}
\sum_\bk\left[  \overline t_{dx\uparrow \downarrow} d^\dagger_{\bk \uparrow} p_{x,\bk\downarrow}
+\overline t_{dx\downarrow \uparrow} d^\dagger_{\bk \downarrow} p_{x,\bk\uparrow} \right ]
\end{equation}
is invariant because first term in the sum transforms into the second,
\begin{eqnarray}
\sum_\bk \Theta^{-1} \overline t_{dx\uparrow \downarrow} d^\dagger_{\bk \uparrow} p_{x,\bk\downarrow} \Theta &=& \sum_\bk \overline t_{dx\uparrow \downarrow}^\ast d^\dagger_{-\bk \downarrow} (-1)^2 p_{x,-\bk\uparrow} \nonumber \\
 &=& \sum_\bk \overline t_{dx\downarrow \uparrow} d^\dagger_{\bk \downarrow}  p_{x,\bk\uparrow},
\end{eqnarray}
and vice versa.  Key to this is that $t_{dx\uparrow\downarrow}$  is an even function of $\bk$.

A similar approach can be taken for rotations.  Under a rotation by $\pi/2$, 
\begin{eqnarray}
R_{\pi/2}^{-1} d_{\bk  s} R_{\pi/2} &=& - e^{i s \pi/4} d_{\overline \bk  s},\\
R_{\pi/2}^{-1}p_{x,\bk s} R_{\pi/2} &=& e^{i s \pi/4} p_{y, \overline \bk  s}, \\
R_{\pi/2}^{-1} p_{y,\bk s} R_{\pi/2} &=& -e^{i s \pi/4} p_{x, \overline \bk  s}. 
\end{eqnarray}
with $\overline \bk = (k_y, -k_x)$.
Under this transformation (making the $k$-dependence of the matrix elements explicit),
\begin{eqnarray}
R_{\pi/2}^{-1}\left[ \sum_\bk \overline t_{dx\uparrow\downarrow}  d^\dagger_{\bk\uparrow} p_{x,\bk\downarrow} \right ] R_{\pi/2}&=& i \sum_\bk \overline t_{dx\uparrow\downarrow}(\bk) d^\dagger_{\overline \bk\uparrow} p_{y, \overline \bk\downarrow} \nonumber\\
&=& \sum_{\overline \bk} \overline t_{dy\uparrow\downarrow}(\overline \bk) d^\dagger_{ \overline \bk\uparrow} p_{y,  \overline \bk\downarrow}\nonumber \\
&=& \sum_{ \bk} \overline t_{dy\uparrow\downarrow}( \bk) d^\dagger_{  \bk\uparrow} p_{y,   \bk\downarrow}. \nonumber \\
\end{eqnarray}
The second line makes use of Eqs.~(\ref{eq:tdxud}) and (\ref{eq:tdyud}).  Overall, this equation is an example of how different matrix elements of Eq.~(\ref{eq:mat6b}) transform into one another under rotation.


%

\end{document}